\documentclass[letter]{jpsj2} 
\usepackage{graphicx, color}
\def\nle{\ \raise.3ex\hbox{$<$}\kern-0.8em\lower.7ex\hbox{$\sim$}\ }
\def\nge{\ \raise.3ex\hbox{$>$}\kern-0.8em\lower.7ex\hbox{$\sim$}\ }
\def\ia{{\accent 19 \char 16}}

\title{ Peculiar `{ from-Edge-to-Interior}' Spin Freezing
in a Magnetic Dipolar Cube}

\author{Katsuyoshi \textsc{Matsushita}$^{1}$\thanks{E-mail: kmatsu@issp.u-tokyo.ac.jp},
Ryoko \textsc{Sugano}$^{2}$\thanks{E-mail: sugano@rd.hitachi.co.jp},
Akiyoshi \textsc{Kuroda}$^{1}$\thanks{E-mail: kro@issp.u-tokyo.ac.jp},
Yusuke \textsc{Tomita}$^{1}$\thanks{E-mail: ytomita@issp.u-tokyo.ac.jp} and
Hajime \textsc{Takayama}$^{1}$\thanks{E-mail: takayama@issp.u-tokyo.ac.jp}}
\inst{$^{1}$ Institute for Solid State Physics, University of Tokyo, 5-1-5 Kashiwanoha, Kashiwa, Chiba, 277-8581, Japan.\\
$^{2}$ Advanced Research Laboratory, Hitachi, Ltd.,
   1-280 Higashi-Koigakubo, Kokubunji-shi, Tokyo 185-8601, Japan.}
\recdate{\today}

\abst{
By molecular dynamics simulation, we have investigated classical
Heisenberg spins, which are arrayed on a finite simple cubic lattice and
interact with each other only by the dipole-dipole
interaction{, and have found its peculiar 
{\it from-Edge-to-interior} freezing process. As the temperature is
decreased, spins on each edge predominantly start to freeze in a
ferromagnetic alignment parallel to the edge around the corresponding
bulk transition temperature, then from each edges grow domains with short-range
orders similar to the corresponding bulk orders, and the system ends up
with a unique multi-domain ground state at the lowest temperature. We 
interpret this freezing characteristics} is attributed to the anisotropic
and long-range nature of the dipole-dipole interaction combined with a finite-size
effect. 
}

\kword{Magnetic dipolar system, nanomagnetism, magnetic domain, spin dynamics simulation, spin-freezing characteristics}

\begin{document}
\sloppy
\maketitle

\noindent
{
In the last decade systems consisting of arrayed ferromagnetic
nanoparticles have attracted much attention as a possible element with
the high storage density~\cite{Chou}. Magnetic properties of such
systems have been analyzed based on various theoretical model, in which
the magnetic moment of each nanoparticle is represented by a classical
Heisenberg spin with a proper magnitude. The spins are interacting with
each other by the dipole-dipole interaction, and each spin suffers from
the magnetic anisotropy energy which represents the shape and/or bulk
lattice anisotropies of the original nanoparticle magnetic
moments. There appeared studies on the susceptibility~\cite{Arias},
magnetic hysteresis~\cite{Fuzi,Szabo}, and energy
relaxation~\cite{Fernandez1} in such models of a finite size. However,
there have been little work on systems only with the dipole-dipole
interaction which we call here simply dipolar systems. Considering that
to clarify magnetic properties of dipolar systems of a finite size is of
importance as the first step in researches of arrayed magnetic
nanoparticles as well as of interest from a theoretical viewpoint in
statistical physics, we address this problem in the present work.}

\begin{figure}[bt]
\begin{center}
\includegraphics[width=0.6\linewidth]{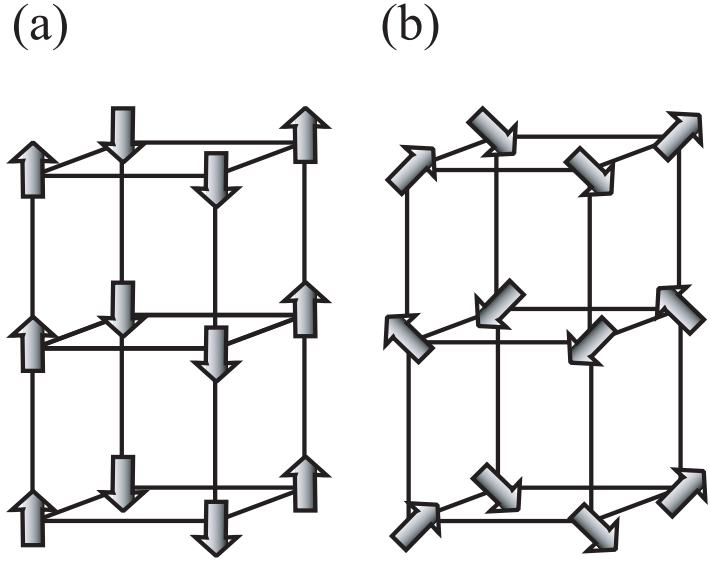}
\end{center}
\caption{The spin configurations in the ground state.  (a)
 Antiferromagnetically aligned ferromagnetic chains (af-FMC) order, and
 (b) one of the LT  orders which have the continuous $O(3)$ symmetry.}\label{fig0}
\end{figure}

The ground-state magnetic structure of the { bulk} dipolar
system was studied more than a half century ago~\cite{Sauer,Luttinger}. 
In particular, Luttinger and Tisza (LT)~\cite{Luttinger} derived
analytically that of cubic lattices, restricting themselves to magnetic
superlattice structures of 2$\times$2$\times$2 unit. {The
LT ground state on the simple cubic (SC) lattice is given by any
properly-normalized linear combination of the three basic states 
(Fig.~\ref{fig0}(a)), one of which is shown in Fig.~\ref{fig0}(b): spins
on $z$-axis align ferromagnetically along the axis and these
ferromagnetic chains align antiferromagnetically in the $x$ and $y$
directions (the rest two are defined by changing the directions
$x,y,z$).  We call them the antiferromagnetically aligned ferromagnetic
chains (af-FMC) orders.} 
Surprisingly, a very limited number of works on the model have been done
since then~\cite{Belobrov,Romano1}. Recently we have checked the LT
ground states by the Monte Calro (MC) analysis~\cite{Tomita}. We have
found the lower energy states with the longer magnetic superlattice
structures than those of the LT ones for the body centered cubic
lattice, while for the SC lattice no lower energy state than the LT one
has been found.   

In the present work we focus on a dipolar cube cut out from the SC
dipolar lattice, with its edges being parallel to the lattice axes, and
look for finite-size effects on its ground-state structures and its
freezing characteristics by the stochastic molecular dynamics (MD)
simulation. We have found very peculiar aspects which one cannot think
of for finite-size systems with any short-range interaction. The ground
state of the dipolar cube consists of 12 domains, each of which
{is very close to} one of three af-FMC orders mentioned
above. A very peculiar aspect we have observed in the cooling process is
what we call {\it from-edge-to-interior} freezing 
characteristics: spins on the edges first freeze independently from
spin freezing on other edges around the transition
temperature of the corresponding bulk SC dipolar system, the af-FMC
domains grow from the edges to the interior at lower temperatures, and
then finally a unique combination of the domains is reached at the
lowest temperature.  

The model we study consists of classical Heisenberg spins, ${\vec S}$'s, 
which are arrayed on a finite cube. {They correspond to
magnetic} moments of single ferromagnetic domain particles normalized by
their magnetization $M_{\rm s}$. The dipole-dipole interaction energy
between the spins is written as 
\begin{eqnarray}
{\cal H} = \frac{J}{2a^3}\sum_{ r_{ij}} {\vec S}_i \cdot \frac{1-{\vec e}_{ij} \otimes {\vec e}_{ij} }{r_{ij}^3} \cdot {\vec S}_j. \label{Ham}
\end{eqnarray}
Here ${r}_{ij}$ denotes dimensionless length (normalized by the lattice
constant, $a$) between sites $i$ and $j$, and ${\vec e}_{ij}$ the unit
vector along the direction from site $i$ to site $j$. To the coupling
constant, $J$, the value of $M_s^2$ is contained, and $J/a^3$ is put the
energy (and temperature { $T$} with $k_{\rm B}=1$)
unit of the present model. For example, in Ni nanoparticles array with
30 nm particle diameter and $a\simeq 100$~nm,$^{3)}$ $J/a^3$ is estimated
about 5 K. If Co is replaced for Ni, $J/a^3$ becomes of order of several
tens K. It is also worth quoting a typical value of magnetic molecular crystal, $J/a^3 \simeq$ 0.7 K for $S_z$ = 10 with $a$ $\simeq$ 0.7 nm \cite{Friedman,Wernsdorfer}. 

The freezing characteristics of the model is analyzed by the
Landau-Lifshiz-Gilbert (LLG) eq. \cite{LLG},  
\begin{eqnarray}
\frac{d{\vec S_i}}{dt} = \frac{\gamma}{1+\alpha^2} {\vec S_i} \times \left\{ {\vec H_{{\rm eff},i}} - \frac{\alpha}{M_s} {\vec S_i} \times {\vec H_{{\rm eff},i}} \right\}, \label{LLG}
\end{eqnarray}
with the effective magnetic field, ${\vec H_{{\rm eff},i}}$, given by
\begin{eqnarray}
{\vec H_{{\rm eff},i}} = \frac{\partial{\cal H}}{\partial\vec{S}_i} + {\vec f}_i, 
\end{eqnarray} 
where ${\cal H}$ is the dipole-dipole interaction of Eq.~(\ref{Ham}).
{To include the heat-bath effect of temperature $T$
we introduce random force ${\vec f}_i$, its root mean square value being}
proportional to $T\alpha$. $\alpha$ and $\gamma$ are so-called
Gilbert damping constant and the gyromagnetic constant, respectively.  In the present study we solve the above
set of equations based on the Euler method with conserving the norm of
spins ($M_s = 1$) and by setting $\alpha a^3/M_s \gamma J$ as the time unit.  We further
put {both constants $\alpha$ and $\gamma$ to 1.0, and the
time step of integrating Eq.~(\ref{LLG}), $\Delta t$, to 0.01.}  With
this choice of the parameter values, the Larmor precession of a spin due  
to the internal field of averaged magnitude damps in a time comparable
with one period of the precession at $T=0$. 

In the present work we mainly report the results obtained by simulations
on a cube whose linear size, $L$, is set to 16. It is an optimal size to
look for peculiar phenomena in a magnetic dipolar cube: the size should
be large enough to enable us to observe typical behavior of spins both
on the boundary (corners, edges, and surfaces) and inside the
cube, while simulation on a too large system is beyond computational
power available for us.  We note that, in the present simulation
the sum in Eq.~(\ref{Ham}) are faithfully taken into account except for
the analysis to obtain the result shown in Fig.~\ref{fig4}(b) below.

We perform typically 8 cooling runs with a fixed cooling rate, in which
we use different sets of random numbers generating their initial
configurations as well as fluctuation forces. The average over them is
carried out to obtain nearly-equilibrium, or frozen values 
of the observables reported below unless otherwise indicated. In each
cooling run, temperature $T$ is initially set to 1 ($=J/a^3$) and is 
decreased by a step of $\Delta T = 0.02$. 
{
At each temperature Eq.~(\ref{LLG}) is integrated by a period of} 
$\tau=$ 5 $\times$ 10$^4$$\Delta t$, which corresponds to about 500 periods of
the Larmor precession mensioned above. {The} average value
of quantities of our interest at each temperature is defined by that
over the whole period of $\tau$.  For example, the freezing of
individual spins, ${\cal S}_i$, is evaluated as
\begin{eqnarray}
 {\cal S}_i = \frac{1}{\tau} \left|\int_\tau \vec{S}_i(t) dt\right|.\label{Si}
\end{eqnarray}

\begin{figure}[bt]
\begin{center}
\includegraphics[width=\linewidth]{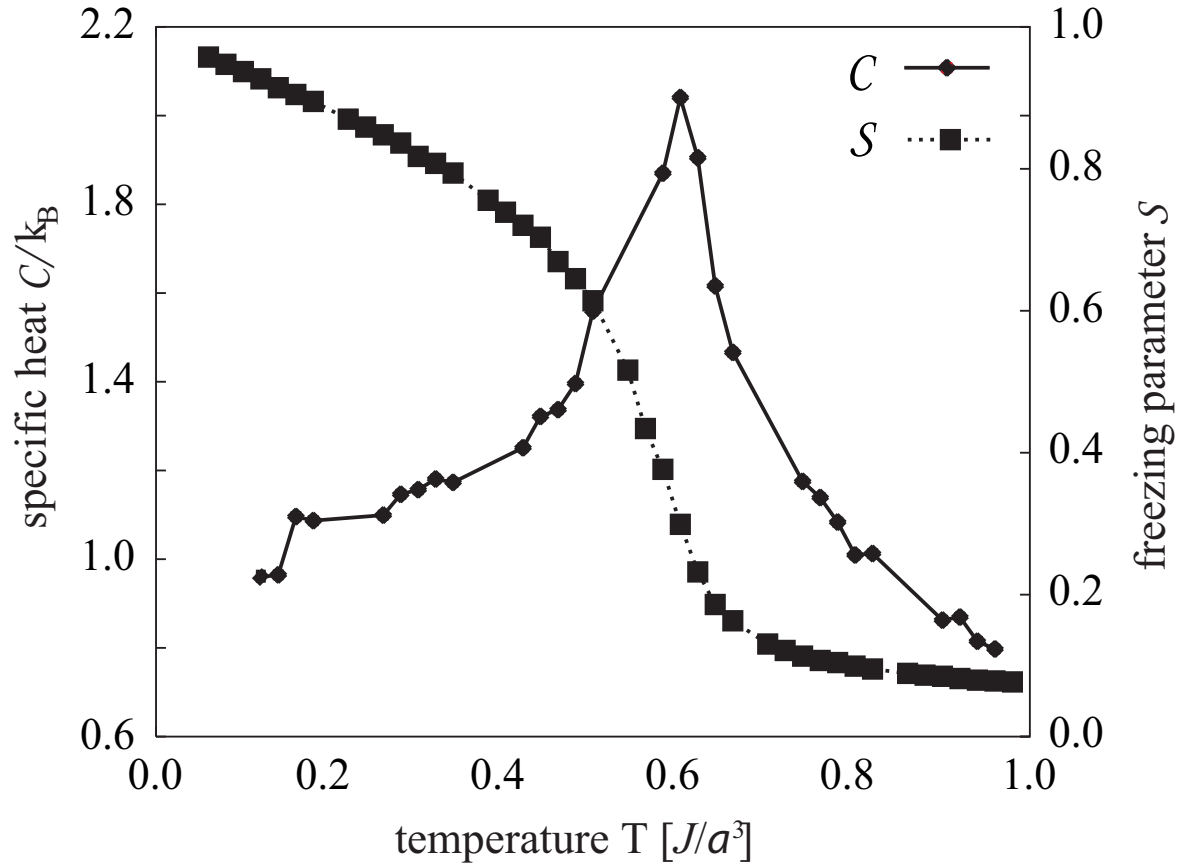}
\end{center}
\caption{The specific heat ${\cal C}$ and the freezing parameter ${\cal S}$  as a function of temperature T.}\label{fig2}
\end{figure}

Let us first present the results of global thermodynamic quantities. In
Fig.~\ref{fig2}, we demonstrate the specific heat, ${\cal C}$, and the
overall freezing parameter, ${\cal S}$, as a function of { 
temperature.} Here ${\cal S}$ is defined as 
${\cal S} = N^{-1}\sum_i {\cal S}_i$, including the average over the
cooling runs, where $N$ is the total number of spins in the cube.
It represents on average how much each spin is frozen in a time scale of
$\tau$. We see from Fig.~\ref{fig2} that the specific heat
exhibits a peak at $T^* \sim 0.6$, which is closed to the transition
temperature, $T_{\rm c}=0.56$, evaluated previously in the corresponding
bulk system~\cite{Romano1}. 
Correspondingly, ${\cal S}$ starts to rapidly increase around $T^*$. 
Both ${\cal C}$ and ${\cal S}$ clearly indicate occurrence of cooperative
freezing around $T^*$.

\begin{figure}[h]
\begin{center}
\includegraphics[width=\linewidth]{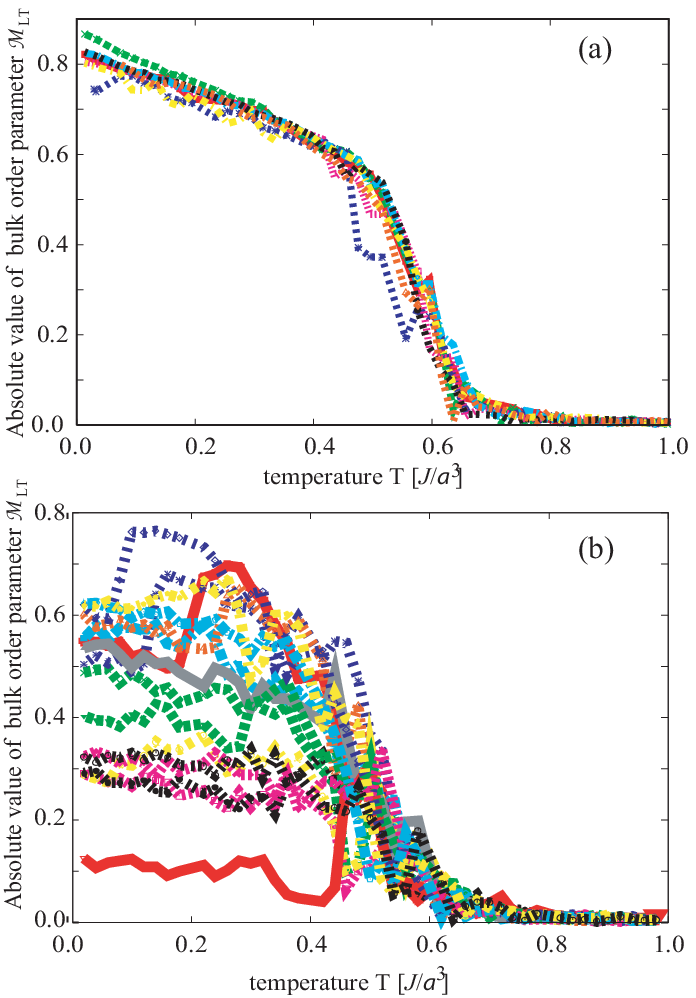}
\end{center}
\caption{The LT order parameter ${\cal M}_{\rm LT}$ of the dipolar cube
 observed in 8 freezing runs; a) without and b) with cut-off for the
 dipole-dipole interaction.}\label{fig4}
\end{figure}

It is natural for us to expect that the frozen order we found in the
cube somehow reflects the magnetic structure of the bulk ground-state
derived analytically by LT. {The order parameter, which
represents the LT ground-state order (see Fig.~\ref{fig0}(a,b)),} is
defined by a vector,  
$\vec{\cal M}_{\rm LT}$, whose $\alpha$-component is written 
{as}
\begin{eqnarray}
 {\cal M^{\alpha}}_{\rm LT} = \frac{1}{\tau} \int_{\tau} \sum_i (-1)^{\sum_{\beta \ne \alpha} n_i^\beta} S_i^\alpha(t) dt,\label{ST} 
\end{eqnarray}
where $\alpha, \beta \in \{x,y,z\}$, and $n_i^\alpha$ denotes the
$\alpha$-coordinate of site $i$. The absolute magnitudes of 
$\vec{\cal M}_{\rm LT}$, denoted as ${\cal M}_{\rm LT}$, for the 8
cooling  runs are shown in Fig.~\ref{fig4}(a). 
All ${\cal M}_{\rm LT}$'s start to rapidly increase from 0 around 
$T^*$ with decreasing temperature and reach almost a unique value of about 0.8
which is {significantly} less than the value of the bulk
order, 1.0, at the lowest temperature. One plausible interpretation of
the result is that spins on the cube reach {a stable state
which consists} of the domains with different af-FMC orders.
{Figure~\ref{fig4}(b), on the other hand, are 
${\cal M}_{\rm LT}$'s obtained by the identical simulation as above but
with one difference, namely, the dipole-dipole interaction is cut off
within the range of $L/2$ in this simulation. By the introduction of the
cut-off, the freezing behavior of ${\cal M}_{\rm LT}$ fluctuates
drastically among 8 cooling runs, though ${\cal C}$ and ${\cal S}$ do
not exhibit such a qualitative difference (not shown).}

\begin{figure}[tb]
\begin{center}
\includegraphics*[width=\linewidth]{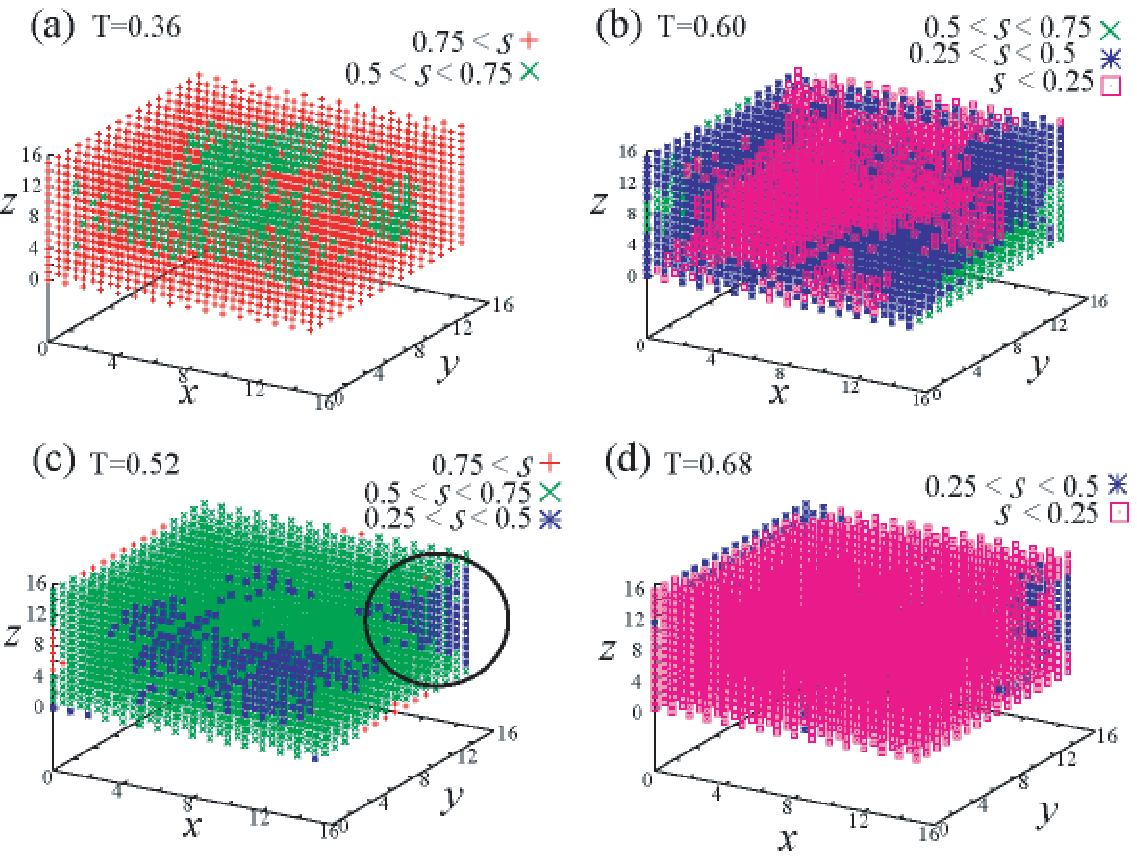}
\end{center}
\caption{Patterns of $\{S_i\}$'s defined by eq.~\ref{Si} for T=0.625, 0.5, 0.375 and 0.125.}\label{fig3}
\end{figure}

To check the above interpretation {on the result shown in 
Fig.~\ref{fig4}(a)}, we examine freezing patterns $\{{\cal S}_i\}$
defined by Eq.~(\ref{Si}) of {whole the cube. 
Figures}~\ref{fig3}(a-d) are typical result observed in one of the
cooling processes. At $T=0.68\ (\geq T^*)$,  spins on some
edges start to freeze. As the temperature further decreases ($T = 0.6 \leq T^*$), 
the freezing extends to the whole boundary in the sense that boundary
spins are relatively less fluctuating than inner spins.  This feature is
kept at further lower temperatures. At $T=0.52$, however, spins in a
domain indicated by the circle in Fig.~\ref{fig3}(c) look as if they are
partially melting. Finally at very low temperatures such as $T=0.36$,
also inner spins almost freeze in the time scale of our observation. We
can say that, in a cooling process of the present dipolar cube, spins on
edges are the most stable, and those on boundary surfaces are stabler
than those inside the cube. 

\begin{figure}[b]
\begin{center}
\includegraphics[width=\linewidth]{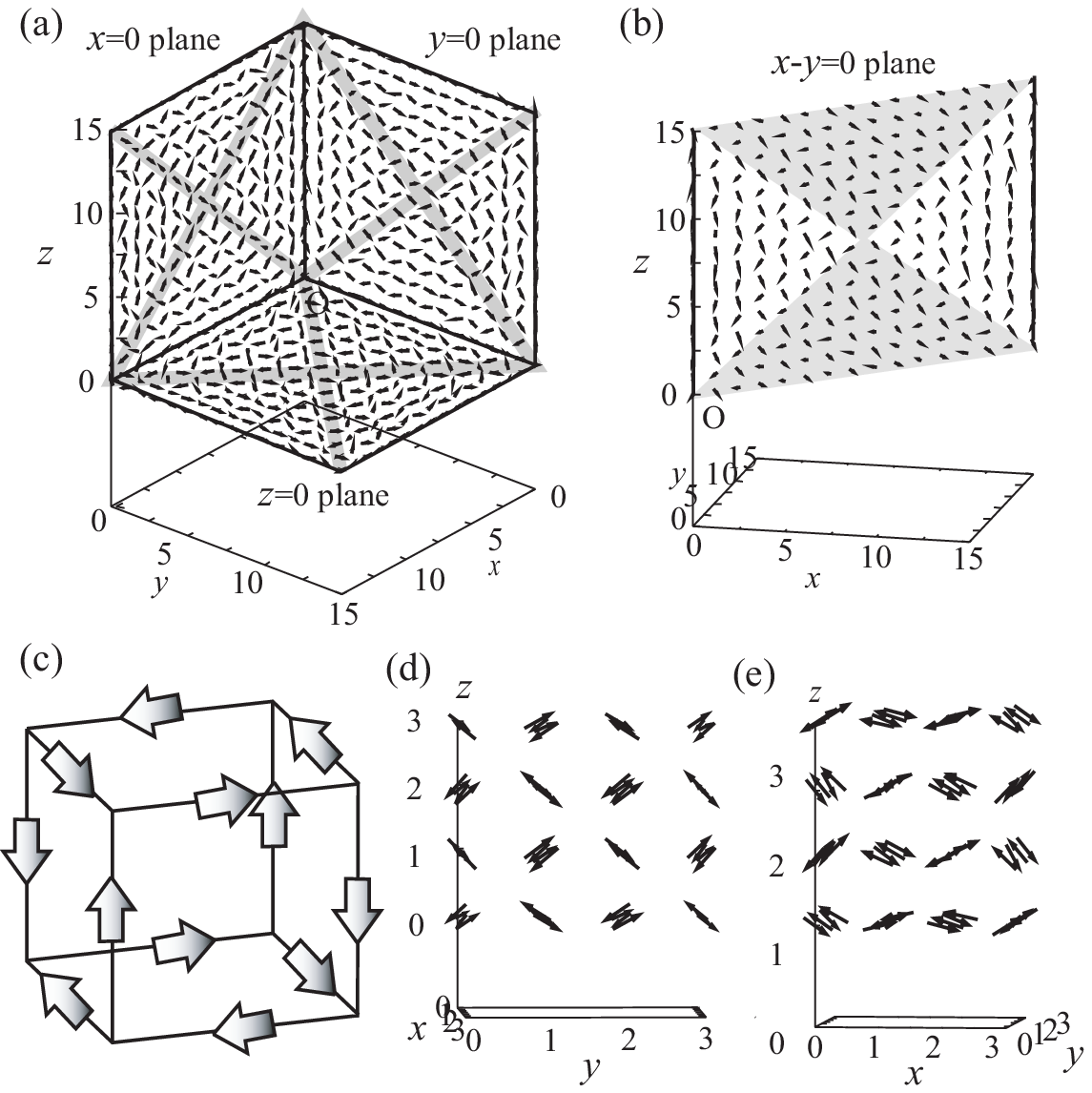}
\end{center}
\caption{The spin configurations for the surfaces (a) and interior (b)
 of the dipolar cube at lowest temperatures. The shaded regions denote
 domain walls.  Figure (c) represents the domain order in the ground
 state of the dipolar cube. Figures (d,e) are the spin configuration in the
 ground state of the $L$=4 dipolar cube. }\label{fig1}
\end{figure}

A typical spin configuration at a very low temperature {
reached after the cooling process is shown in Figs}.~\ref{fig1}(a,b). A
common property shared with such spin configurations is that spins on a
boundary surface (including an edge) tend to align in the direction
parallel to the surface (the edge).  This tendency can be naively
attributed to the anisotropic nature of the dipole-dipole interaction of
Eq.~(\ref{Ham}). Namely, on a surface where there are two (one) axes
parallel (perpendicular) to the surface, a ferromagnetic chain (FMC)
aligned in parallel to the surface is more favorable. Naturally, on an
edge an FMC order parallel to the edge is most favorable. Consequently,
we can also expect that the magnetic structure at the lowest temperatures
consists  of 12 domains with short-range order which is {
close to} one of the three af-FMC orders whose FMC order develops along
the edge of each domain.  This is in fact seen in
Figs.~\ref{fig1}(a,b). Also, if the simulation is started from one of
the af-FMC configurations such as show in Fig.~\ref{fig0}(a), it ends up
with a similar configuration to the one shown in 
Figs.~\ref{fig1}(a, b){: the spins} on a boundary initially
in the direction perpendicular to the boundary flop into the parallel
directions to the boundary.   

By further inspection of the spin configurations, we find that the one
with the lowest energy satisfies the following two conditions; the FMC
orders on three edges which cross at each corner take
2-in(out)-1-out(in) configuration and those on opposite sides of a
surface is aligned antiparallelly. These two conditions make the
{combination of domains in the ground state} to be unique
one which is shown in Fig~\ref{fig1}(c){. Here} structures
which can be connected by $\pi$/2-rotation around 4-fold axis of the
cubic sample are regarded as identical. {In this context,
we want to emphasize that the proper self-rearrangement of the domains}
is due to the very long-range nature of the dipole-dipole interaction. 
In fact, {when the cut-off of the interaction is
introduced, each cooling run ends up with states with different values
of ${\cal M}_{\rm LT}$ as indicated in Fig.~\ref{fig4}(b), i.e., the
adjustment of the af-FMC domains to reach their ground-state combination
hardly takes place when the} dipole-dipole interaction is cut off.  

{One more important freezing aspect we have observed is as
follows.} In the ground-state configuration, {individual}
local energies of the corner spins are higher than those of other spins
as expected. However, those of the edge spins are not necessarily lower
than those of the interior spins. This implies that the relatively
stronger freezing tendency discussed above can {not be
simply attributed to energetics of individual spins. Instead, we have to
consider} cooperative nature of the freezing. As noted above, the bulk
LT order has the continuous $O(3)$ symmetry, and so different LT
short-range orders are expected to grow in the bulk system, or inside
the cube as well, at higher temperatures than $T_{\rm c}$ or 
$T^*\ (\simeq T_{\rm c})$. On the boundary of the cube, however, a
symmetry breaking due to the presence of boundary partially solves the
degeneracy. Especially on an edge, an FMC order along the edge
direction, which has only the $Z_2$ symmetry as a whole, becomes
preferable. This symmetry lowering of the short-range {
order is considered to be an origin of the {\it from-edge-to-interior}
freezing characteristics we have found at and below $T^*$.} Thus the
boundary geometry, combined with the anisotropic nature of the
dipole-dipole interaction, also plays an important role on the 
freezing {nature of the dipolar cube}. 

So far the dipolar cube with $L=16$ has been discussed. For an odd-$L$
dipolar cube, { one of the two conditions for the
ground-state configuration for an even-$L$ cube mentioned above is
certainly violated, since FMC's on opposite sides of an odd-$L$ square in
its ground state are checked} to be parallel to each other.\cite{Sugano} 
We also note the stable configurations in cubes with
small $L$'s. { In the $L=16$ cube, spins around a vertex of
the cube take an optimized {direction, which} is regarded as
a part of the so-called vortex structures.~\cite{Belobrov}} In small
cubes with $L=4,6,8$, this corner structure extends to {a
size comparable with $L/2$, and consequently their ground state consists
of a single-domain vortex state as shown in Fig.~\ref{fig4}(d,e) which
differs distinctly from the configuration shown in Figs.~\ref{fig1}(a,b).}
{We can expect variety of stable configurations of SC dipolar
cubes arising from the
relative angles between edges of the cube and the lattice axes. Further
variety of magnetic properties of finite dipolar systems is expected
when we think of the shape effect, different lattice structures, and so
on. The present work has just opened a door in this interesting field.} 

To conclude let us summarize a scenario of the 
{\it from-edge-to-interior} freezing process we have found in the
dipolar cube.  1) At
higher temperatures than $T^*\ (\simeq T_{\rm c})$, the LT short-range
orders are expected to grow.
2) Around $T^*$, where
some of the LT short-range orders become in touch with one of the cube
edges, the degeneracy in the LT orders is lifted, and one of the af-FMC
orders fitted to  the edge direction becomes stabilized. 3) The
domains with the af-FMC order extends from edge to interior at lower
temperatures. 4) In case the combination of domain structures does not
match with the one of the ground-state, upturns of certain domains
occur, and 5) the spin configuration reaches to the unique
ground-state at lowest temperatures. 

The present work is supported by NAREGI Nanoscience Project from the
Ministry  of Education, Culture, Sports, Science, and Technology. The
numerical simulations have been partially performed also by using the 
facilities at the Supercomputer Center, Institute for Solid State
Physics, the University of Tokyo.

\end{document}